\documentclass{eptcs}
\providecommand{\event}{TFPIE 2017} 
\usepackage{breakurl}             
\usepackage[noxcolor]{pstricks}
\usepackage{pst-node}
\usepackage{graphicx}
\usepackage{subfigure}
\usepackage{stfloats}
\usepackage{multido}
\usepackage{setspace}
\usepackage{textcomp}
\usepackage{amssymb}
\usepackage{alltt}

\title{Vector Programming Using Structural Recursion \\ \Large{An Introduction to Vectors for Beginners}}
\author{Marco T. Moraz\'an
\institute{Seton Hall University}
\email{morazanm@shu.edu}}
\def\titlerunning{Vector Programming Using Structural Recursion}
\def\authorrunning{M. T. Moraz\'an}
\begin{document}
\maketitle

\begin{abstract}
Vector programming is an important topic in many Introduction to Computer Science courses. Despite the importance of vectors, learning vector programming is a source for frustration to  many students given that they feel left adrift when it comes to resolving vector indexing errors. Even though the size of a vector is a natural number, there have been no efforts to define a useful recursive data definition to help beginners design vector processing functions. This article defines the concept of a vector interval and describes how to exploit its recursive structure to design vector processing functions. The described methodology provides a context beginners can use to reason about proper vector indexing instead of leaving them adrift with this responsibility. A key feature of properly using the described methodology is that if students process the correct vector interval then vector indexing errors can not arise. The classroom deployment of this approach is described in detail. Students, to date, have found vector intervals helpful in avoiding out-of-bounds indexing errors when all the vector elements of the interval are processed.
\end{abstract}

\section{Introduction}
It may very well be true that every college-trained computer programmer remembers long nights debugging programs that manipulate vectors. Perhaps, many readers of this article recall hours of work trying to determine why an index into a vector was out of range. In some programming languages, this is equivalent to figuring out why a program caused a core dump. If these same readers ponder about this long enough, they may recall that many of those long frustrating nights occurred when they were first exposed to vectors and beginning to learn how to program. Surprisingly, the same holds true for many undergraduates today. Frankly, it is a bit shocking how little we have advanced as a community in teaching vector programming to beginners.

Vector programming (a.k.a. array programming), of course, is extremely important in Computer Science. Vectors, given that they are random access, allow us to efficiently implement algorithms for applications in a myriad of fields. Vectors, for example, allow us to efficiently implement various data structures such as binary trees \cite{Ford}, stacks \cite{Goodrich}, and process queues \cite{Silberschatz}. They also allow us to reduce the complexity of algorithms such as, for example, finding a path in a directed graph \cite{HtDP} and matrix multiplication \cite{Burden}. Vectors are also useful in the reduction of memory allocation by, for example, allowing us to sort, say, files or numbers in place \cite{Knuth3}. Given the importance and versatility of vectors it is in our interest as a community to make an introduction to vector programming for beginners as frustrationless as possible. This can be achieved by providing beginners with a structured model they can use to reason about processing vectors.

The crux of the problem with much of the developed introductory material to vector programming is that it provides a structureless definition of a vector. Some material goes a step further and presents an \textsf{ADT} for vectors. Some of these \textsf{ADT}s describe a single operation: indexing. It is, thus, not surprising that many beginners feel as being left adrift to figure out on their own how to avoid out-of-bounds indexing errors. This is highly undesirable, in the experience of the author, for at least two reasons. The first is that students get frustrated enough to quit Computer Science as a major. This is an especially important issue in universities, like in the USA, where beginning students spend a semester or two shopping around for a major. The second is that students that persevere  develop bad programming habits associated with the belief that vector programming is a strict exercise in trial and error instead of an exercise in design.

This article describes the work developed to introduce students to vector programming at Seton Hall University (SHU). At the heart of the approach is providing students with data definitions that help them design vector processing functions to solve problems. As the length of a vector can be of arbitrary size, these must be recursive data definitions. Such data definitions can be directly exploited using structural recursion to develop programs. The model presented to beginning students is that of a \emph{vector interval} that may only contain valid indices into a vector and is used to process all the elements in the interval. Thus, reducing frustration over out-of-bounds errors. The article is organized as follows. Section \ref{rw} discusses related work. Section \ref{sb} discusses the programming background of the students with whom this approach is used. Section \ref{intervals} introduces and defines the concept of an interval. In addition, it provides examples of how to design functions using the different data definitions for an interval. Section \ref{vp} shows how students are introduced to vector programming by developing data definitions for a vector interval. Section \ref{examples} discusses three extended vector programming examples using the described approach. Finally, Section \ref{conclusions} presents concluding remarks and directions for future work.

\section{Related Work}
\label{rw}
Many textbooks introduce vectors as lacking a recursive structure that can be exploited to solve problems. Readers are then introduced to how to use vectors using snippets of code that emphasize that an index into the vector must be within its bounds. For example, a vector is described as a collection of variables of the same type with each element having an index \cite{Goodrich} or as a finite sequential list of elements of the same datatype identifying the first element, the second element, the third element, and so forth \cite{Ford}. Such data definitions are inadequate, because they hide the recursive nature of the interval of valid indices into the vector and focus exclusively on the syntax to declare, create, index, and mutate vectors. They fail to provide the proper model to help beginners design functions/methods that process a vector avoiding, for example, illegal indexes into the vector. The second data definition may even be considered misleading by describing a vector as a list that has a well-known recursive structure. Vectors do not have a decomposable structure like lists. Furthermore, describing vector elements as the first, the second, the third, and so forth does not assist in any way the design of vector processing functions. We can not program ``and so forth." Even some modern approaches to make programming popular among the young address vectors in a very similar manner (e.g., \cite{codeorg}). In contrast, the work presented in this article aims to provide students with a decomposable data definition that beginners can use to reason about vector processing. This data definition is that of a vector interval. These intervals have a recursive structure that guides the design of vector processing functions.

The problem of only using legal indices into a vector is traditionally and summarily left to the student with no clear indication of how to accomplish this task (e.g., \cite{Sedgewick}). Some may convincingly argue that this is relatively easy when you need to process an entire vector. However, the matter is not clear when you need to process only part of a vector. Consider sorting a vector using insertion sorting which requires inserting elements in the sorted part of a vector. It is difficult for a beginner to determine or be confident that she is always correctly indexing the vector given that the size of the sorted part of the vector is not constant. Even worse, however, is that it is more difficult to pin down bugs when indexing errors occur if a model that helps the student reason about indexes is not provided to them. In contrast, the work presented in this article helps students reason about the indexes into a vector when all elements within an interval are processed. A vector interval only spans the valid indexes of the part of the vector being processed. As such, when the structure of an interval is used to design a function students know that any natural number in the vector interval is a valid index. If the vector interval is empty then students know that the vector should not be indexed. 

Some efforts have gone beyond syntax. The textbook \emph{How to Design Programs} (\textsf{HtDP}), for example, describes a vector as a well-defined mathematical class of data with some basic constructors and observers \cite{HtDP}. \textsf{HtDP} further states that we can think of vectors as functions on a small finite range of natural numbers. This begins to provide some context for vector processing, but surprisingly falls short of identifying the recursive structure of this range of natural numbers as it does so well for other types of data. It is unlikely that future editions of \textsf{HtDP} are to develop this given that the second edition has eliminated its introduction to vector programming \cite{HtDP2}. In contrast, the work presented in this article tackles the recursive nature of this finite range of natural numbers and describes it as an interval. Furthermore, the fact that this range is finite is carried to its logical conclusion to obtain two data definitions for a vector interval. One data definition leads to a template that is used to design functions to process vectors from right to left (i.e., from the largest index down to the lowest index in the vector interval) and the other is used to design functions that process vectors from left to right (i.e., from the lowest index to the largest index in the vector interval).

Related to the material discussed in this article, although not directly, are ubiquitous efforts to hide low-level vector manipulation from programmers. For example, an \textsf{Iterator} generates a sequence of elements from a collection, one at a time \cite{Tymann}. This allows programmers to examine every element of a vector without worrying about properly indexing the vector. The details of properly indexing the vector are hidden by the \textsf{Iterator}. Similarly in functional programming, for example in \textsf{Racket}, programmers have access to functions like \textsf{vector-map} and \textsf{vector-filter} that iterate through a vector and that hide the details of properly indexing the vector. Clearly, iterators and higher-order vector functions process the entire interval of valid indices into a vector. Programming with vector intervals using structural recursion, as presented in this article, also disallows improperly indexing a vector. In contrast, however, these vector intervals are not restricted to only the entire range of valid indices into a vector. A vector interval can be used to process any continuous subinterval of valid indices. Furthermore, whether in a functional or object-oriented language, iterators and higher-order vector functions must be implemented in the first place and this is certainly a context in which reasoning about vector intervals is valuable.

Vector intervals, in this article, are defined in terms of the size, \textsf{N}, of a vector. The idea is that a valid interval of indices into a vector must be in [0..N-1]. This notion is similar to that found in dependently typed programming languages such as Dependent ML \cite{Xi} and Idris \cite{Brady}. In Dependent ML, for example, \textbf{array} is used as a built-in type constructor that takes as input a type for the elements of the array and a natural number, \textsf{n}, for the size of the array. The array can only be indexed if the index is a natural number less than \textsf{n}. This notion is not enforced by vector intervals, but vector intervals do begin to plant the notion in students that data definitions can depend on dynamic properties of the data being processed.

\section{Student Background}
\label{sb}
At SHU, the introductory Computer Science courses span two semesters and focus on problem solving using a computer \cite{mtm22,mtm24}. The languages of instruction are the successively richer subsets of \textsf{Racket} known as the student languages which are tightly-coupled with \textsf{HtDP} \cite{HtDP,HtDP2}. No prior experience with programming is assumed. Before introducing students to vector programming, the course familiarizes students with primitive data (e.g., numbers, strings, booleans, symbols, and images), primitive functions, and library functions to manipulate images (i.e., the image teachpack). During this introduction, students are taught about variables, defining their own functions, and the importance of writing contracts and purpose statements. The next step of the course introduces students to data analysis and programming with compound data of finite size (i.e., structures). At this point, students are introduced to the first design recipe. Students develop experience in developing data definitions, examples for data definitions, function templates, and tests for all the functions they write. A great deal of emphasis is placed on all of these steps as part of the problem-solving design process. Building on this experience, students develop expertise on processing compound data of arbitrary size such as lists, natural numbers, and trees. In this part of the course, students learn to design functions using structural recursion. After structural recursion, students are introduced to functional abstraction and the use of higher-order functions such and \textsf{map} and \textsf{filter}. The first course ends with a module on distributed programming \cite{mtm26}.

In the second course students are exposed to generative recursion, accumulative recursion, and mutation \cite{mtm25}. The course starts with generative recursion. At the end of this module, students get their first exposure to vector programming. Students are taught the syntax needed for vectors and are introduced to the design of vector processing functions using the material on intervals and vector intervals outlined in this article.  After this, the course exposes students to accumulative recursion and iteration. The course ends with two modules on mutation that include their second exposure to vector programming. In this second exposure, students design vector mutators using vector intervals.

The topics covered follow much of the structure of \textsf{HtDP} \cite{HtDP}. There are two 75-minute lectures every week and the typical classroom has between 20 to 25 students. In addition to the lectures, the instructor is available to students during office hours (3 hours/week) and there are 20-30 hours of tutoring each week. Students may voluntarily attend any number of tutoring hours they like. The tutoring hours are conducted by undergraduate students handpicked and trained by the author. These tutors focus on making sure students develop answers for each step of the design recipe (from writing contracts to running tests). Students must attempt to follow the steps of the design recipe prior to attending tutoring. Based on a student's work, the tutors provide guidance but do not solve problems. Students are still responsible for successfully completing all steps of the design recipe. In addition, tutors attend lectures to assist students when they get stuck with, for example, syntax errors. This type of team-teaching with undergraduate tutors has proven to be extremely well-received by students and to be an effective means to enhance the learning experience.

\section{Intervals}
\label{intervals}
Before introducing students to vectors, they are re-introduced to the concept of an interval. The term re-introduced is used in the same manner as students being re-introduced to natural numbers earlier in the course. That is, students in general are familiar with at least one of the following ``definitions" for the set of natural numbers:
\begin{alltt}
          \(\mathbb{N} = \{0,1,2,3,\ldots\} \)     \(\mathbb{N} = \{1,2,3,\ldots\}\).
\end{alltt}
Both of these definitions are inadequate, because they do not describe how to construct a natural number. Furthermore, students are left to figure out the meaning of $\ldots$ . Knowing how to construct a natural number is important, because it empowers students with the knowledge needed to process such numbers by exploiting their structure. Therefore, a more useful data definition for the set of natural numbers is (e.g., adopted in \textsf{HtDP}):
\begin{alltt}
          A natural number (natnum) is either:
               1. 0
               2. (add1 n), where n is a natural number.
\end{alltt}
Such a definition exposes the structure of natural numbers and is used to define a template to write functions that process a natural number:
\begin{alltt}
          f-on-natnum: natnum \(\rightarrow\) \(\ldots\)
          Purpose: \(\ldots\)
          (define (f-on-natnum n)
            (cond [(= n 0) \(\ldots\)]
                  [else n\(\ldots\)(f-on-natnum (sub1 n))])).
\end{alltt}
The body of this template, in essence, states that the conditional distinguishes between the varieties of natural numbers. For each variety an expression is needed to compute the result. When a natural number, \textsf{n}, is a constructed natural number (i.e., the second variety) the expression can manipulate the value of \textsf{n} and can recursively process, \textsf{(sub1 n)}, the natural number used to construct \textsf{n}. This template is then specialized by students every time they need to solve a problem that requires processing a natural number. Specializing, in this context, means filling in the blanks (i.e., the different $\ldots$).

Students bring to the classroom an understanding about intervals analogous to their initial understanding of natural numbers. That is, they define an interval as:
\begin{alltt}
          [\(i..j\)], where \(i \leq j\)
\end{alltt}
Once again, such a definition is inadequate. It does not expose the structure of an interval that is helpful to solve problems that require processing an interval. Furthermore, the fact that an interval can be empty is well-hidden by such a definition. Given that students are already familiar with recursive data definitions, it is not much of an intellectual leap to re-define an interval, \textsf{INTV}, as:
\begin{alltt}
     An interval, INTV, is two integers, low and high, such that it
     is either:
         1. empty, if low > high
         2. [[low..(sub1 high)]..high], where [low..(sub1 high)] is an INTV
                                        and low \(\leq\) high
\end{alltt}

The natural way to represent intervals is with a structure or an object that has two fields. Choosing a two-integer representation is a concession to current practices in existing programming textbooks. To the best knowledge of the author, there are no programming books that capture in a structure or an object the lowest and highest indices of an interval. A judgement call had to be made between representing an interval as two integers or as a structure/object. Given that beginning students will read programming books that explicitly use two indices to process a vector, the two-integer representation was ultimately chosen. It does have the advantage that it makes the material feel somewhat familiar to students that arrive in the classroom with vector programming experience.

The \textsf{INTV} data definition makes the structure of an interval explicit. Students know that given that there is variety in the data definition a conditional is needed to distinguish among the different varieties. Furthermore, students can observe that when the interval is not empty \textsf{high} is a whole number constructed using \textsf{n}. This means that \textsf{[low..(sub1 high)]} is part of the structure of \textsf{[low..high]}. Put differently, \textsf{[low..(sub1 high)]} is used to construct \textsf{[low..high]}. For example, the interval \textsf{[-1..1]} is constructed as follows:
\begin{alltt}
     [-1..1] = [[-1..0]..1]
             = [[-1..-1]..0..1]
             = [[-1..-2]..-1..0..1]
             = [empty..-1..0..1]
\end{alltt}
Now it becomes clear that when the interval is not empty the value of \textsf{high} as well as the result of recursively processing the subinterval \textsf{[low..(sub1 high)]} can be used. This naturally leads to the following function template to process an interval:
\begin{alltt}
     ; f-on-INTV: int int \(\rightarrow\) \(\ldots\)
     ; Purpose: For the given INTV, \(\ldots\)
     (define (f-on-INTV low high)
       (cond [(empty-INTV? low high) \(\ldots\)]
             [else high\(\ldots\)(f-on-INTV low (sub1 high))]))
\end{alltt}
This template requires a function to detect that an interval is empty. For the chosen representation using two integers, this function and tests are easily developed by students:
\begin{alltt}
     ; empty-INTV?: int int \(\rightarrow\) boolean
     ; Purpose: For the given INTV, determine if it is empty
     (define (empty-INTV? low high) (> low high))

     (check-expect (empty-INTV? 3 4) false)
     (check-expect (empty-INTV? 30 30) false)
     (check-expect (empty-INTV? 5 4) true)
\end{alltt}

The template can now be used to solve problems that process an interval. For instance, consider computing the summation of all the integers in an interval. Students know to start with the template for an \textsf{INTV} and to develop an answer for each variety of the data starting with the non-recursive case. Students quickly observe that when the interval is empty the summation is 0. When the interval is not empty, they observe that \textsf{high} must be added to the result of recursively processing \textsf{[low..(sub1 high)]}. Observe how reasoning about the structure of an interval leads the programmer to a solution. Putting these ideas together leads to the following specialization of the template and tests for summing the elements of an interval:
\begin{alltt}
     ; sum-INTV: int int \(\rightarrow\) int
     ; Purpose: For the given INTV, sum its elements
     (define (sum-interval low high)
       (cond [(empty-INTV? low high) 0]
             [else (+ high (sum-INTV low (sub1 high)))]))

     (check-expect (sum-INTV 10 1) 0)
     (check-expect (sum-INTV 10 10) 10)
     (check-expect (sum-INTV -1 1) 0)
\end{alltt}

After working out some exercises, students realize that the template suggests that intervals must be processed from \textsf{high} to \textsf{low} (or right to left). However, many students also realize that it may be equally correct to process an interval from \textsf{low} to \textsf{high} (or left to right). This requires the development of the following data definition for an interval:\\
\begin{alltt}
      An interval2, INTV2, is two integers, low and high, such that it
      is either:
          1. empty, if low > high
          2. [low..[(add1 low)..high]], where [(add1 low)..high] is an INTV2
                                        and low <= high
\end{alltt}
In this data definition, \textsf{low} is constructed by subtracting 1 from some integer \textsf{n}. This leads to the following function template:
\begin{alltt}
     ; f-on-INTV2: int int \(\rightarrow\) \(\ldots\)
     ; Purpose: \(\ldots\)
     (define (f-on-INTV2 low high)
       (cond [(empty-INTV2? low high) \(\ldots\)]
             [else low\(\ldots\)(f-on-INTV2 (add1 low) high)]))
\end{alltt}
It is important to highlight to students that the above template is not an instance of generative recursion. Many students see \textsf{add1} and associate it with generative recursion and mistakenly feel they must develop a termination argument for functions written using this template.

Once armed with this knowledge, students can now solve problems processing the interval from left to right. For instance, summing all the integers in an interval can also be solved as follows:
\begin{alltt}
     ; sum-INTV2: int int \(\rightarrow\) int
     ; Purpose: For the given INTV, sum its elements
     (define (sum-INTV2 low high)
       (cond [(empty-INTV2? low high) 0]
             [else (+ low (sum-INTV2 (add1 low) high))]))

     (check-expect (sum-INTV2 10 1) 0)
     (check-expect (sum-INTV2 10 10) 10)
     (check-expect (sum-INTV2 -1 1) 0)
\end{alltt}

\section{Vector Processing}
\label{vp}
Armed with an understanding of how to process intervals, students are ready to be introduced to vectors. After introducing students to what a vector is, why it is desirable to use them, and some basic vector-syntax, students are explained that it is common to process a contiguous subset of a vector. The emphasis here is on \emph{common} given that arrays are random access and can be processed in many different ways. Nonetheless, the reader  is reminded that the goal is to expose students for the first time to vectors and, as such, intervals are useful to reason about and design programs to process a contiguous subset of a vector.

For example, for a given vector \textsf{V},  we may want to process the entire vector (from  indices 0  to \textsf{(sub1 (vector-length V))} or we may want to process only part of the vector (from indices \textsf{a} to \textsf{b}). Clearly, processing a contiguous subset of a vector requires processing an interval. Once again, this is not a huge intellectual leap for students. Care must be taken, however, because we must avoid attempting to access a vector with an illegal index that is either negative or greater than or equal to the length of the vector. This requires developing a data definition for a  \emph{vector interval}. A vector interval is an interval that places restrictions on what values \textsf{low} and \textsf{high} may take. In general, a valid index into a vector, \textsf{V}, is between 0 and \textsf{(sub1 (vector-length V))}. Thus, we can define a vector interval as follows:
\begin{alltt}
     For a vector of size N, a vector interval, VINTV, is two integers,
     low \(\geq\) 0 and -1 \(\leq\) high \(\leq\) N-1, such that it is either:
       1. empty  (i.e., low > high)
       2. [[low..(sub1 high)]..high], where [low..(sub1 high)] is a VINTV
                                      and low \(\leq\) high
\end{alltt}
Observe that this data definition restricts a \textsf{VINTV} to only contain valid indices into the vector when it is not empty. That is, it depends on \textsf{N}. The indices are all natural numbers. Further observe that the structure of a vector interval is exactly the same as the structure of an interval. There is a difference when processing a \textsf{VINTV}. We are interested in processing vector elements instead of interval elements. This means that in the body of a function to process a \textsf{VINTV} a vector must be referenced. As with intervals, a data definition that leads to processing a vector interval from left to right is also developed.

\begin{figure}[t]
\begin{alltt}
     ; f-on-vector: (vector X) \(\rightarrow\) \(\ldots\)
     ; Purpose: \(\ldots\)
     (define (f-on-vector V)
       (local [; f-on-VINTV: int int \(\rightarrow\) \(\ldots\)
               ; Purpose: For the given VINTV, \(\ldots\)
               (define (f-on-VINTV low high)
                 (cond [(empty-VINTV? low high) \(\ldots\)]
                       [else (vector-ref V high)
                             \(\ldots\)(f-on-VINTV low (sub1 high))]))

               ; f-on-VINTV2: int int \(\rightarrow\) \(\ldots\)
               ; Purpose: For the given VINTV2, \(\ldots\)
               (define (f-on-VINTV2 low high)
                 (cond [(empty-VINTV2? low high) \(\ldots\)]
                       [else (vector-ref V low)
                             \(\ldots\)(f-on-VINTV2 (add1 low) high)]))]
         \(\ldots\)))
\end{alltt}
\caption{The Template for Functions on Vectors.}
\label{vtemplate}
\end{figure}

The above observations allow for the in-class development of the function template to process a vector displayed in Figure \ref{vtemplate}. The contract states that any function that processes a vector must take as input at least a vector of any type (\textsf{X} is a type variable). The body of the function is a \textsf{local}-expression that may be used to define one or more local functions and values. Students are told that problem analysis will reveal the type of expression that is needed in the body of the \textsf{local}-expression. If a single value is needed from the given vector, then the expression will be one that processes a vector interval. Otherwise, the expression will be one that uses different values obtained from processing the same vector. The local definition section contains two templates: one for each direction that a vector interval can be processed in. At least one of the templates is to be used to process vector elements. Observe that in each of the local templates, vector elements are processed (using \textsf{vector-ref}) instead of  interval elements.

\begin{figure}[t]
\begin{alltt}
     ; avg-vector: (vectorof number) \(\rightarrow\) number
     ; Purpose: To compute the average of the given vector
     ; Assumption: The vector is not empty.
     (define (avg-vector V)
       (local [; sum-elems: int int \(\rightarrow\) natnum
               ; Purpose: For the given interval, sum the
                          vector elements
               (define (sum-elems low high)
                 (cond [(empty-VINTV? low high) 0]
                       [else (+ (vector-ref V high)
                                (sum-elems low (sub1 high)))]))]
         (/ (sum-elems 0 (sub1 (vector-length V)))
            (vector-length V))))

     (check-within (avg-vector (vector 6 7 8 9)) 7.5 0.01)
     (check-within (avg-vector (vector 1 2 3)) 2 0.01)
\end{alltt}
\caption{A Function to Compute the Average of a Vector of Numbers.}
\label{avgv}
\end{figure}

To make the use of the function template to process a vector concrete, consider computing the average of a vector of numbers. Problem analysis reveals that the vector cannot be empty given that division by 0 is undefined. It also reveals that it does not matter in which direction the \textsf{VINTV} is processed as addition is a commutative operation. Now, the template for functions on a vector from Figure \ref{vtemplate} is specialized. The contract indicates that the input is a vector of numbers, \textsf{V}, and that the function returns a number. The body of the \textsf{local}-expression must be an expression that divides the sum of the vector elements by the length of the vector. This means that we must write a function to compute the sum of vector elements. Given our problem analysis, either of the templates to process a vector interval can be used. Without loss of generality, we can choose to process from right to left (i.e., the template for \textsf{VINTV}). This means that when the vector interval is empty the answer is 0 and that when it is not empty we add \textsf{(vector-ref V high)} to the result of recursively processing the rest of the \textsf{VINT} (i.e., \textsf{[low..high-1]}). The resulting function and its tests are displayed in Figure \ref{avgv}. Observe that by using the template based on structural recursion it is impossible to have indexing errors if a valid vector interval is provided as the initial input to \textsf{sum-elems}. Thus, simplifying the job for beginners.

\section{Extended Examples}
\label{examples}
This section presents three extended examples of how to design functions that process vectors. The first, the dot product of two vectors \cite{Baase,Venit}, is an example of how to process multiple vector intervals simultaneously in step. The second, the merging of two sorted vectors \cite{Knuth3}, is an example of how to process multiple vector intervals that are not processed in step. This example also shows that the design of vector mutators can benefit from exploiting the structure of vector intervals. The third, insertion-sort \cite{HtDP,Knuth3}, sorts a vector in place. This is an example of how reasoning about vector intervals, not vector indexing, assists in the design of functions/mutators.

\subsection{The Dot Product of Two Vectors of Numbers}
Given two vectors of numbers, \textsf{V1} and \textsf{V2}, the dot product is defined as:
\begin{alltt}
     V1 \(\cdot\) V2 = \(\Sigma\sb{i=0}\sp{N}\) V1[i]*V2[i], where N is the length of the vectors minus 1.
\end{alltt}
In-classroom problem analysis reveals:
\begin{itemize}
  \item Both vectors must have the same length.
  \item Both vectors must be entirely processed simultaneously in step.
  \item Vector elements can be processed either right to left or vice versa.
\end{itemize}
Given these insights, students conclude that the function can be designed around processing a single vector interval, say for \textsf{V1}, and then specialize the template for functions on vectors to develop their code. The input is two vectors of numbers and the output is a number. This is reflected in the contract in Figure \ref{dotv}. The body of the \textsf{local}-expression calls a function, \textsf{sum-products}, to process the single interval that spans all the elements of both vectors (i.e., from 0 to \textsf{(sub1 (vector-length V1))}). Observe that the interval given to \textsf{sum-products} only contains valid indices into the vector. Thus, indexing errors cannot occur when using the function template.

The function, \textsf{sum-products}, is designed using either template for vector interval processing given the third insight above. Figure \ref{dotv} displays tests and the result of using the template that processes the vector interval from left to right. The code is developed by steps, as before, by formulating answers for each variety of vector interval. Students have no trouble seeing that the answer is 0 when the interval is empty. When the interval is not empty, students are explained that they must do something with the two elements, in this case, at the low end of each vector. This is what it means to process both vectors simultaneously in step. This action, of course, is to multiply them. To formulate the final answer, this product must be added to the result of recursively processing the rest of the vector interval.

\begin{figure}[t]
\begin{alltt}
; dot-product: (vector number) (vectorof number) \(\rightarrow\) number
; Purpose: To compute the dot product of the two given vectors
(define (dot-product V1 V2)
  (local [; sum-products: int int \(\rightarrow\) number
          ; Purpose: For the given VINTV, compute the dot product of V1 and V2
          (define (sum-products low high)
            (cond [(empty-VINTV2? low high) 0]
                  [else (+ (* (vector-ref V1 low) (vector-ref V2 low))
                           (sum-products (add1 low) high))]))]
    (sum-products 0 (sub1 (vector-length V1)))))

(check-within (dot-product (vector) (vector)) 0 0.01)
(check-within (dot-product (vector 1 2 3) (vector 1 2 3)) 14 0.01)
\end{alltt}
\caption{Function to Compute the Dot Product of Two Vectors.}
\label{dotv}
\end{figure}

As stated, no indexing errors can arise in this example. The key to success is for students to properly define the initial \textsf{VINTV2} to be processed. In  this case, this task is fairly straightforward given that both vectors must be processed  in their entirety. Also observe that there is no guess work involved in how to process the rest of the elements in each vector. The solution to that concern is baked into the template for functions on a vector.

\subsection{Merge Two Sorted Vectors}

\begin{figure}[t]
\begin{alltt}
     ; merge: (vector number) (vectorof number) \(\rightarrow\) (vectorof number)
     ; Purpose: To merge the two given sorted vectors in non-decreasing order
     ; Assumption: The given vectors are sorted in non-decreasing order
     (define (merge V1 V2)
       (local [; res: (vectorof number)
               ; Purpose: To store the merged elements so far
               (define res (build-vector (+ (vector-length V1)
                                            (vector-length V2))
                                         (lambda (i)  (void))))

               ; combine: int int int int int int \(\rightarrow\) (vectorof number)
               ; Purpose: For the given VINTVs, merge V1 and V2 into res
               (define (combine lowv1 highv1 lowv2 highv2 lowres highres)
                 \(\ldots\))
       (combine 0 (sub1 (vector-length V1))
                0 (sub1 (vector-length V2))
                0 (sub1 (vector-length res)))))

     (check-expect (merge (vector) (vector)) (vector))
     (check-expect (merge (vector 10) (vector 2)) (vector 2 10))
     (check-expect (merge (vector 1 4 6) (vector 2 4 5 8 9))
                   (vector 1 2 4 4 5 6 8 9))
\end{alltt}
\caption{Basic Outline for a Function to Merge Two Sorted Vectors.}
\label{mergev}
\end{figure}

Consider the problem of merging two vectors that are sorted in non-decreasing order into a single vector that is sorted in non-decreasing order. In-classroom problem analysis yields the following insights:
\begin{itemize}
  \item A vector to hold all the elements of the given vectors must be allocated. Given that this vector must be mutated every time an element is added, it must be a state variable.
  \item Three different intervals must be processed simultaneously: one for each of the input vectors and one for the result vector. These intervals are not processed in step.
  \item Each vector must be processed in its entirety.
\end{itemize}
Figure \ref{mergev} displays the basic outline for a function to merge two sorted vectors obtained from beginning to specialize the template for a function on a vector. The contract, the purpose statement, and the assumption indicate that two sorted vectors of numbers are expected as input and a sorted vector is expected  as output. The body of the \textsf{local}-expression calls an auxiliary function, \textsf{combine}, to process the three vector intervals. Three intervals are needed as input,  because they are not processed in step. Observe that all three initial vector intervals span all the valid indices, respectively, for each vector. Thus, by using structural recursion indexing errors cannot occur. Locally, the state variable, \textsf{res}, is defined to store the result. Its invariant states that it is a vector of numbers. This  vector is initialized to contain only \textsf{void} values to indicate that nothing in the vector has been initialized. The tests demonstrate the expected behavior with vectors of varying sizes.

\begin{figure}[t]
\begin{alltt}
     (cond [(and (empty-VINTV2? lowv1 highv1)
                 (empty-VINTV2? lowv2 highv2))
            res]
           [(empty-VINTV2? lowv1 highv1)
             (begin
               (vector-set! res lowres (vector-ref V2 lowv2))
               (combine lowv1 highv1
                        (add1 lowv2) highv2
                        (add1 lowres) highres))]
           [(empty-VINTV2? lowv2 highv2)
            (begin
              (vector-set! res lowres (vector-ref V1 lowv1))
              (combine (add1 lowv1) highv1
                       lowv2 highv2
                       (add1 lowres) highres))]
           [(< (vector-ref V1 lowv1) (vector-ref V2 lowv2))
            (begin
              (vector-set! res lowres (vector-ref V1 lowv1))
              (combine (add1 lowv1) highv1
                       lowv2 highv2
                       (add1 lowres) highres))]
            [else
              (begin
                (vector-set! res lowres (vector-ref V2 lowv2))
                (combine lowv1 highv1
                         (add1 lowv2) highv2
                         (add1 lowres) highres))]))]
\end{alltt}
\caption{The Conditional for the function \textsf{combine} from Figure \ref{mergev}.}.
\label{condmerge}
\end{figure}

The task left is to develop the body of \textsf{combine}. Given that three intervals are not processed in step, we need to determine the different conditions that may arise during processing to augment the \textsf{cond}-expression that appears in the template to process a vector interval. After some class discussion, the conclusion is reached that at each step an element of one of the input vectors is placed in the result vector. Furthermore, the input vectors and the result vector are to be processed from left to right. This is an implementation choice and it is equally correct to process the intervals from right to left. These new insights and our implementation choice, in conjunction with the previous insights, lead us to five cases that must be addressed:
\begin{enumerate}
  \item The intervals for both input vectors are empty.
  \item The interval for the first input vector is empty.
  \item The interval for the second input vector is empty.
  \item The low element of the first input vector is less than the low element of the second input vector.
  \item The low element of the second input vector is less than or equal to the low element of the first input vector.
\end{enumerate}
Observe that the non-recursive case is listed first and must be the first to be solved. For this case, there are no more elements to process in either vector interval for the input vectors and the result vector is returned. For the second case, the vector interval for the first vector is empty and the process continues by placing the remaining elements left in the second vector interval into the result vector. The recursive call is made with the rest of the vector intervals for both the second input vector and the result vector. The third case is the same as the second case, but it is the vector interval for the second vector that is empty. The recursive call is made with the rest of the vector intervals for both the first input vector and the result vector. The fourth and fifth cases place the smallest \textsf{low} element of the input vectors into the result vector. The recursive call is always made with the rest of the interval for the result and the rest of the interval for the input vector that had an element placed in the result. The resulting conditional  is displayed in Figure \ref{condmerge}. Once again, observe that the proper use of structural recursion eliminates the possibility of indexing errors.

The important lesson to derive from this example is that when more than one interval is processed then every recursive call must be made with the rest of the intervals that process either the \textsf{low} or the \textsf{high} vector element. The key to avoiding indexing errors is now reduced to, once again, making sure the initial vector intervals are valid.

\subsection{Insertion-sort In Place}

Insertion-sort is an algorithm that may be used to sort a vector in place. Sorting a vector, \textsf{V}, for a given vector interval is summarized as follows:
\begin{itemize}
  \item If the vector interval is empty, stop.
  \item If the vector interval is not empty
   \begin{itemize}
     \item Sort the rest of the interval.
     \item Insert the first element of the vector interval into the sorted subinterval.
   \end{itemize}
\end{itemize}

\begin{figure}[t]
\begin{alltt}
; insort-in-place!: (vector number) \(\rightarrow\) (void)
; Purpose: To sort the given vector in non-decreasing order
; Effect: Rearrange the elements of the given vector in non-decreasing order
(define (insort-in-place! V)
  (local [; insert!: int int \(\rightarrow\) (void)
          ; Purpose: For the given VINTV2, insert V[low] in V[low+1..high]
          ;          such that V[low..high] is in non-decreasing order
          ; Effect: V elements are swapped until one is >= V[low] or
          ;         the given VINTV2 is empty
          (define (insert! low high)
            (cond [(empty-VINTV? low high) (void)]
                  [else (cond [(<= (vector-ref V low)
                                   (vector-ref V (add1 low))) (void)]
                              [else (begin (swap low (add1 low))
                                           (insert! (add1 low) high))])]))
          ; sort!: int int \(\rightarrow\) (void)
          ; Purpose: For the given VINTV2, sort V using insertion sort
          ; Effect: For the given VINTV2, rearrange V in non-decreasing order
          (define (sort! low high)
            (cond [(empty-VINTV2? low high) (void)]
                  [else
                   (begin (sort! (add1 low) high)
                          \((insert! low high)\))]))]
    (sort! 0 (sub1 (vector-length V)))))
(define VINS0 (vector 10))
(define VINS1 (vector 10 3 7 17 11))
(check-expect (begin (insort-in-place! VINS0) VINS0) (vector 10))
(check-expect (begin (insort-in-place! VINS1) VINS1) (vector 3 7 10 11 17))
\end{alltt}
\caption{Buggy Insertionsort in Place.}
\label{insv}
\end{figure}

A basic outline for the insertionsort function is displayed in Figure \ref{insv}. Readers familiar with \textsf{HtDP}, will recognize the basic outline. Here its presentation has been adapted to use the concept of a vector interval. The code displayed is typical of what is developed in the classroom by student-led discussion using the template for a function on a vector. The function takes as input a vector of numbers and returns \textsf{void} as it is a mutator. The entire vector must be sorted and this is reflected by the vector interval given as input to the auxiliary function \textsf{sort!}. The tests illustrate that both the desired effect/mutation is achieved and the desired result is returned.

Class discussion leads to having \textsf{sort!} process the vector interval from left to right. Thus, the template to process a \textsf{VINTV2} is used. It is made clear, however, that the vector interval can be processed from right to left and this is left as a homework exercise. Students quickly realize that if the given \textsf{VINTV2} is empty then the mutator should stop and this is reflected by returning \textsf{void}. If the given \textsf{VINTV2} is not empty, then the rest of the vector interval is sorted. This part is not controversial for students. After sorting the rest of the given \textsf{VINTV2}, students typically state that the vector interval \textsf{[low..high]} must be processed again to place the \textsf{low} element in the right place.

The function \textsf{insert!} places the \textsf{low} element of the given \textsf{VINTV2} in the right place. Once again, students perform the required case analysis on a \textsf{VINTV2}. If the vector interval is empty, there is nothing to place in the right place and the mutator stops returning \textsf{void}. Is the given \textsf{VINTV2} is not empty, the \textsf{low} and \textsf{low + 1} elements are tested. If they are not out of order, the mutator stops returning \textsf{void} as \textsf{V} is sorted for the given vector interval. If they are out of order, the mutator swaps them and proceeds to complete the inserting operation using the rest of the given \textsf{VINTV2}.

\begin{figure}[t]
\begin{alltt}
; insort-in-place!: (vector number) \(\rightarrow\) (void)
; Purpose: To sort the given vector in non-decreasing order
; Effect: Rearrange the elements of the given vector in non-decreasing order
(define (insort-in-place! V)
  (local [; insert!: int int \(\rightarrow\) (void)
          ; Purpose: For the given VINTV2, insert V[low] in V[low+1..high]
          ;          such that V[low..high] is in non-decreasing order
          ; Effect: V elements are swapped until one is >= V[low] or
          ;         the given VINTV2 is empty
          ; Assumptions: high+1 is a valid index into V
          ;              low \(\neq\) high
          (define (insert! low high)
            (cond [(empty-VINTV? low high) (void)]
                  [else (cond [(<= (vector-ref V low)
                                   (vector-ref V (add1 low))) (void)]
                              [else (begin (swap low (add1 low))
                                           (insert! (add1 low) high))])]))
          ; sort!: int int \(\rightarrow\) (void)
          ; Purpose: For the given VINTV2, sort V using insertion sort
          ; Effect: For the given VINTV2, rearrange V in non-decreasing order
          (define (sort! low high)
            (cond [(empty-VINTV2? low high) (void)]
                  [else (begin (sort! (add1 low) high)
                               (insert! low (sub1 high)))]))]
    (sort! 0 (sub1 (vector-length V)))))
(define VINS0 (vector 10))
(define VINS1 (vector 10 3 7 17 11))
(check-expect (begin (insort-in-place! VINS0) VINS0) (vector 10))
(check-expect (begin (insort-in-place! VINS1) VINS1) (vector 3 7 10 11 17))
\end{alltt}
\caption{Insertionsort in Place.}
\label{ins}
\end{figure}

Much to their surprise, students get an out of bounds error when they run their tests. The yet unknown bug manifests itself in the \textsf{insert!} function when comparing the \textsf{low} and \textsf{low + 1} elements. Students observe that an attempt is made to index the vector at \textsf{(vector-length V) + 1} and are now asked how an indexing error can occur if the template to process a vector interval is correctly used. After some vivid class discussion, the students determine the answer. The template has not been correctly used. The bug is that the initial vector interval given to \textsf{insert!} is incorrect (highlighted in italics in Figure \ref{insv}). To correctly insert the \textsf{low} element, \textsf{sort!} must provide \textsf{[low..(sub1 high)]} as the initial vector interval to \textsf{insert!}. Observe that now \textsf{V[low + 1]} can never cause an indexing error in \textsf{insert!}. That is, \textsf{insert!} will never try to index \textsf{V} at \textsf{(vector-length V) + 1}. Furthermore, observe that it is impossible for \textsf{low = high} in \textsf{insert!}. Therefore, we need not worry about \textsf{insert!} not properly working if it were ever called with non-empty vector interval of size 1. The mutator \textsf{insert!} is annotated with assumptions to reflect these observations. The correct code for insertion-sorting in place is displayed in Figure \ref{ins}. The tests for \textsf{insort-in-place!} explicitly include a test for a vector of size 1 to highlight that the mutator works when the interval of valid indices is \textsf{[a..a]} (i.e., \textsf{high = low}).

The important lesson to derive from this example is that beginning students must be led to reason about vector intervals and not about vector indexing errors like done by most, if not all, textbooks. Never mind that interpreters and compilers manifest bugs as vector indexing errors. Unless taught to reason about vector intervals, students spend a frustrating amount of time trying to find an inexistent bug in a function that is correctly written (e.g., the mutator \textsf{insert!} in the example above).

\section{Concluding Remarks}
\label{conclusions}
This article presents a design-oriented methodology to help beginners develop vector processing functions. It is based on the concept of a vector interval that has a recursive structure. This recursive structure is exploited to develop a function template that is specialized by students to solve problems. An important issue that is addressed by this methodology is the proper indexing of vectors. The concept of a vector interval is helpful in avoiding (and resolving) out-of-bounds indexing errors when all the vector elements of the interval are processed. Several examples are presented to illustrate the use of the design methodology in practice. These examples illustrate how to use vector intervals in the context of processing several vectors simultaneously and of finding bugs that manifest themselves as vector indexing errors.

Future work includes formulating generative and accumulative recursion as well as multidimensional-vector processing examples that further demonstrate the usefulness of reasoning about vector intervals to design functions. The work is also being extended to demonstrate how to use vector intervals to reason about algorithms that do not process all the elements of a vector within a given vector interval (e.g., functions used by heapsort). In this case, it is necessary to introduce students to a bit of logic to determine if an index is valid given a vector interval. In addition to functional programming, the work with vector intervals is being extended to a course that focuses on object-oriented design. Finally, on a more long-term basis, students are being surveyed to determine how well vector intervals appeal to them and how useful they find them for vector programming.

\section{Acknowledgements}
Students traditionally believe that the learning process flows from the professor to the students. In my case, nothing can be further from the truth. The work presented in this article is inspired by the difficulties faced and by the questions addressed to me by my beginning students. The author thanks them for providing me with valuable lessons regarding how to teach an introduction to vector programming. In particular, I thank Josephine Des Rosiers for her many heated debates with me about designing programs.

\bibliographystyle{eptcs}
\bibliography{vp}
\end{document}